# Dissipative Quantum Theory: Implications for Quantum Entanglement

A. K. Rajagopal and R. W. Rendell
Naval Research Laboratory, Washington D. C. 20375

Three inter-related topics are discussed here. One, the Lindblad dynamics of quantum dissipative systems; two, quantum entanglement in composite systems and its quantification based on the Tsallis entropy; and three, robustness of entanglement under dissipation. After a brief review of the Lindblad theory of quantum dissipative systems and the idea of quantum entanglement in composite quantum systems illustrated by describing the three particle systems, the behavior of entanglement under the influence of dissipative processes is discussed. These issues are of importance in the discussion of quantum nanometric systems of current research.

I. INTRODUCTION

A consistent description of quantum dissipative systems is provided by the Lindblad equation [1, 2]. Most obviously, dissipative processes occur due to interaction between the system under investigation and its surroundings. The usual von Neumann equation for the density matrix is a unitary description without dissipation. Attempts to incorporate the interaction with the surroundings within this description in the past failed to preserve the basic features of quantum theory, namely the probabilistic nature of the density matrix. The Lindblad equation is a general mathematical formalism which is local in time as in the conventional formulation of quantum theory and preserves all the properties of the density matrix: (a) reality of physical quantities (hermiticity), (b) preservation of basic positive probability measure, (positive semi-definiteness), and (c) conservation of probability (traceclass). Because of its generality, this equation is considered "canonical" in the sense that any derivation of an equation proposing to describe dissipative dynamics, must have the Lindblad form. Recently, under certain assumptions, such an equation for a composite system has been derived[3]. Earlier attempts employing perturbation methods led to an approximate equation which is not in the Lindblad form and violated one or the other of the requirements stated above.

Foremost among problems in composite systems are the questions of entanglement and their importance in realistic physical situations; their persistence and control is crucial in the presence of dissipative effects due to environment etc. This is particularly central in the considerations of nano-devices in the presence of other systems in which they are imbedded. Entanglement appears to be a natural candidate for non-additive description unlike the classical additive descriptions of such systems. It is here that another important aspect of this composite system arises in the development of information theory associated with the density matrix. In this connection, the role of Tsallis entropy and its ramifications for composite systems have become important [4, 5, 6].

In this paper, after a brief description of Lindblad dynamics is given in Sec. II, the notion of entanglement with examples from three qubits is given in Sec.III. Brief introduction to quantum information theory of non-additive systems is also given here. In Sec.IV, control of entanglement etc. is discussed briefly when we incorporate the Lindblad dynamics. We draw on our own recent work in this area [7, 8] in our presentation. We end the paper with a few concluding remarks in Sec. V

II. THE LINDBLAD DYNAMICS

The Lindblad evolution for the density matrix is given by [1]

$$i\partial_t \hat{\rho} = [H, \hat{\rho}] - \frac{i}{2} \sum_{m,n} h_{nm} \left( \hat{L}_m \hat{L}_n^+ \hat{\rho} + \hat{\rho} \hat{L}_m \hat{L}_n^+ - 2 \hat{L}_n^+ \hat{\rho} \hat{L}_m \right) \qquad (1)$$

where $H$ is the Hermitian Hamiltonian operator, $h_{nm}$'s are elements of a c-number Hermitian matrix, $h_{nm}^* = h_{mn}$, and $\hat{L}_m$'s are traceless operators and $\hat{L}_n^+$ is the hermitian conjugate of $\hat{L}_n$. Often, an alternative but equivalent form of eq.(1) is used in the literature which also is useful in establishing certain general properties of the Lindblad equation. This alternative form is obtained by expressing the c-number hermitian coefficient in terms of its eigen expansion:

$h_{nm} = \sum_\lambda h_\lambda C_{\lambda n}^* C_{\lambda m}$, with $h_\lambda$ real. Then we introduce new operators defined by

$\hat{Q}_\lambda = \sum_m C_{\lambda m} \hat{L}_m$ and their hermitian conjugates. Then eq.(1) takes the alternate form

$$i\partial_t \hat{\rho} = [\hat{H}, \hat{\rho}] - \frac{i}{2} \sum_\lambda h_\lambda \left( Q_\lambda Q_\lambda^+ \hat{\rho} + \hat{\rho} Q_\lambda Q_\lambda^+ - 2 Q_\lambda^+ \hat{\rho} Q_\lambda \right). \qquad (2)$$

The Lindblad equation preserves all the properties of the density matrix: (a) traceclass, (b) hermiticity, and (c) positive semi-definiteness. There are two other additional features of this equation of great importance: (d) evolution of pure state into mixed state and vice versa, and (e) unlike many master equations, the short time evolution does not violate positivity of the density matrix, another facet of property (c) above.

It is important to point out that in [3], starting from a system A in contact with another system B, the Lindblad equation for A is derived under some general assumptions and the constants $h_{nm}$'s in eq.(1) are expressed in terms of the interactions between systems A and B. This is a useful result to note. However the generality of the time evolution of the density matrix equation (1, 2) should be noted wherein the constants $h_{nm}$ appearing may be treated as phenomenological in a context-independent way, not necessarily arising from the system being influenced by another.

We will not give proofs of all these statements but only those for (d, e) because these derivations though important are not often given in the literature.

(d) Evolution of pure state into mixed state and vice versa.

Consider the quantity $Tr\hat{\rho}^q$, for arbitrary positive values of q. By calculating its time derivative using eq.(2) and the cyclic property of trace we obtain the result

$$\partial_t Tr\hat{\rho}^q = -q\sum_\lambda h_\lambda Tr\{\hat{Q}_\lambda \hat{Q}_\lambda^+ \hat{\rho}^q - \hat{Q}_\lambda^+ \hat{\rho}\hat{Q}_\lambda \hat{\rho}^{q-1}\}. \tag{3}$$

Using the eigen-expansion

$$\hat{\rho} = \sum_m P_m(t)|m\rangle\langle m|, \tag{4}$$

we obtain

$$\partial_t\left(\sum_m (P_m(t))^q\right) = -q\sum_{m,n,\lambda} h_\lambda |\langle m|\hat{Q}_\lambda|n\rangle|^2 P_m(t)\{(P_m(t))^{q-1} - (P_n(t))^{q-1}\} \tag{5}$$

For q=1, we obtain the result given in (a), namely conservation of trace (total probability). For q different from 1, we obtain an important result if the Q - operators are hermitian. Upon interchanging summation over m,n and adding to the original expression (5), we obtain the following expression:

$$\partial_t\left(\sum_m (P_m(t))^q\right) =$$
$$-\frac{q}{2}\sum_{m,n,\lambda} h_\lambda \left|\langle m|\hat{Q}_\lambda|n\rangle\right|^2 (P_m(t)-P_n(t))\{(P_m(t))^{q-1}-(P_n(t))^{q-1}\} \quad (6)$$

Taking q=2, we can check for the purity of the state. Thus we obtain

$$\partial_t\left(\sum_m (P_m(t))^2\right) = -\sum_{m,n,\lambda} h_\lambda \left|\langle m|\hat{Q}_\lambda|n\rangle\right|^2 (P_m(t)-P_n(t))^2 \leq 0 \quad (7)$$

leading to the result that $\left(\sum_m (P_m(t))^2\right) \leq \left(\sum_m (P_m(t=0))^2\right)$. If initially the system is a pure state ($\hat{\rho}^2 = \hat{\rho}$), the right hand side of the inequality is 1 and so at later times, $Tr\hat{\rho}^2(t)$ will be less than or equal to 1. Hence for finite times, the system evolves into a mixed state ($\hat{\rho}^2 < \hat{\rho}$) from a pure state ($\hat{\rho}^2 = \hat{\rho}$).

(e) Short time evolution:

One of the problems faced with many master equations for the density matrix is that they are frequently found to violate its positive definiteness for short times. This is often proved by integrating the evolution equation for short times close to the given initial time to leading order in perturbation theory as follows: For short times, eq.(2) has a solution in the form

$$\hat{\rho}(\Delta t) = \hat{\rho}(t=0) - i(\Delta t)[\hat{H},\hat{\rho}(0)]$$
$$-\frac{(\Delta t)}{2}\sum_\lambda h_\lambda \left(Q_\lambda Q_\lambda^+ \hat{\rho}(0) + \hat{\rho}(0) Q_\lambda Q_\lambda^+ - 2 Q_\lambda^+ \hat{\rho}(0) Q_\lambda\right) \quad (8)$$

Let us consider an initial solution to be a pure state, $\hat{\rho}(0) = |\Psi_o\rangle\langle\Psi_o|$ constructed from the state of the Hamiltonian, so that the second term is zero. Take the diagonal matrix element of the resulting equation for the density matrix with respect to a state orthogonal to the ground state to find

$$\langle \Psi_1 | \hat{\rho}(\Delta t) | \Psi_1 \rangle = (\Delta t) \sum_\lambda h_\lambda |\langle \Psi_0 | Q_\lambda | \Psi_1 \rangle|^2. \tag{9}$$

For this to be positive requires $h_{nm}$ matrix be positive, which is often violated in the Master-type equations.

We may make one final remark arising from the discussion in (d) above. Let us calculate the time derivative of the Tsallis entropy,

$$S_q(t) = -\left\{ \sum_m (P_m(t))^q - 1 \right\} / (q-1), \tag{10}$$

from the expression obtained in (d):

$$\partial_t S_q(t) =$$
$$\frac{q}{2(q-1)} \sum_{m,n,\lambda} h_\lambda |\langle m | \hat{Q}_\lambda | n \rangle|^2 (P_m(t) - P_n(t)) \left\{ (P_m(t))^{q-1} - (P_n(t))^{q-1} \right\} \tag{11}$$

When q=1, the Tsallis entropy reduces to the von Neumann entropy and this result shows that the von Neumann entropy increases in time whereas the Tsallis entropy increases if q>1 and decreases if q<1. Such a result was proved before for Fokker-Planck evolution whereas here we have shown this for quantum evolution as given by Lindblad (see also [2]). In the next section, we describe quantum composite systems.

III. IDEA OF ENTANGLEMENT: THREE BIT SYSTEMS

To give some feeling for the ideas of entanglement, quantum information theory, etc. we begin with a brief consideration of a composite of two systems A and B, whose density matrix is $\hat{\rho}(A,B)$. We define the marginal density matrix of A as $\hat{\rho}(A) = Tr_B \hat{\rho}(A,B)$ and similarly the marginal density matrix of B. $\hat{\rho}(A,B)$ is said to be entangled if it cannot be put in the form of a complex combination of its marginals: $\sum_i c_i \hat{\rho}_i(A) \otimes \hat{\rho}_i(B)$ where $c_i$ 's are positive constants in [0, 1] whose sum is one. The idea of quantum entanglement is different from the classical entanglement as seen for example, in the bivariate statistical distributions. It should be mentioned that this phenomenon of quantum entanglement is at the heart of very interesting and bizarre features of quantum mechanics, from teleportation to no-exact copying (no-cloning theorem), to quantum computing [9]. The necessary and sufficient conditions for this to happen is not known in general but in special cases of 2x2 and 2x3 cases and in the case of continuous variables, bivariate Gaussians. For bipartite systems, $\hat{\rho}(A,B)$ is a 4x4 matrix and its marginals are 2x2 matrices. One cannot tell by merely examining the density matrix whether it is entangled or not (see however, [10]). A necessary but not sufficient condition based on the conditional Tsallis entropy which is stated in terms of the region where the conditional q-entropy

$$S_q(B|A) = \left(S_q(A,B) - S_q(A)\right) / \left(1 + (1-q)S_q(A)\right) \tag{12}$$

becomes negative. Notice that this procedure needs the entropy of the marginal. This was used in [4] for examining the entanglement status of Werner states defined by

$$\hat{\rho}_W(A,B) = \frac{(1-x)}{4} \hat{I}_2(A) \otimes \hat{I}_2(B) + x |\Psi_2(A,B)\rangle\langle\Psi_2(A,B)| \tag{13}$$

x is a parameter in the range [0, 1] and $\hat{I}_2$ is a 2x2 unit matrix. Here for example, we consider $|\Psi_2(A,B)\rangle = \frac{1}{\sqrt{2}}\left(|A\downarrow\rangle|B\downarrow\rangle + |A\uparrow\rangle|B\uparrow\rangle\right)$. $(\uparrow,\downarrow)$ represent the two states of the qubit in the spin representation. It may be mentioned that in quantum optics, they stand for the horizontal (H) and vertical (V) polarization states of the photon and in

quantum computer parlance, these are represented by (1,0). This is a mixed state density matrix and it is quantum entangled if x>1/3 and classically entangled otherwise. This coincides with other methods proving it to be both necssary and sufficient condition for quantum entanglement. More generally, in [5], similar condition $\left(x > \left(1+N^{n-1}\right)^{-1}\right)$ was found to coincide with the necessary and sufficient condition for entanglement in a NxNx...xN Werner system of an n-partite N - level system.

In [6], a classification of the general pure three qubit states is given based on such criteria along with the permutation symmetry of the states. This is given in Table I. The details of this work may be seen in [6]. The three qubit system described by $\hat{\rho}(A,B,C)$ is interesting in the sense that we have three two-system marginals of the form $\hat{\rho}(A,B) = Tr_C \hat{\rho}(A,B,C)$ and three one-system marginals of the form $\hat{\rho}(A) = Tr_{B,C} \hat{\rho}(A,B,C)$. The Table contains the status of entanglement of pure three particle states as we examine their various marginals. The implication is the manner in which entanglement sustains (Robust, R) or not (Fragile, F) when one loses one and/or two of the states which are represented by the marginals stated above. When the reduction entails less than maximal entanglement as in $|WRr^{\pm}\rangle$, we denote it by r [6]. The three-qubit states considered here are eigenstates of the 3-spin Heisenberg Hamiltonian of the form $\sigma_A \cdot \sigma_B + \sigma_A \cdot \sigma_C + \sigma_B \cdot \sigma_C /2$, with $(|GHZ^{\pm}\rangle, |WRR^{\pm}\rangle)$, $|GFR^{\pm}\rangle$, and $|WRr^{\pm}\rangle$ belonging respectively to the eigenvalues 5/2, -3/2, and -7/2. Explicitly they are given by

$$|GHZ^{\pm}\rangle = \frac{1}{\sqrt{2}}\{|Q_1^+\rangle \pm |Q_1^-\rangle\} \quad |GFR^{\pm}\rangle = |D_2^{\pm}\rangle. \tag{14}$$

$$|WRR^\pm\rangle = |Q_2^\pm\rangle; \quad |WRr^\pm\rangle = |D_1^\pm\rangle. \tag{15}$$

where the 8 three particle states are given by

$$|Q_1^+\rangle = |\uparrow^A \uparrow^B \uparrow^C\rangle; \quad |Q_1^-\rangle = |\downarrow^A \downarrow^B \downarrow^C\rangle;$$

$$|Q_2^+\rangle = \frac{1}{\sqrt{3}}\left\{|\uparrow^A \uparrow^B \downarrow^C\rangle + |\uparrow^A \downarrow^B \uparrow^C\rangle + |\downarrow^A \uparrow^B \uparrow^C\rangle\right\} \tag{16}$$

$$|Q_2^-\rangle = \frac{1}{\sqrt{3}}\left\{|\downarrow^A \downarrow^B \uparrow^C\rangle + |\downarrow^A \uparrow^B \downarrow^C\rangle + |\uparrow^A \downarrow^B \downarrow^C\rangle\right\}$$

$$|D_1^+\rangle = \frac{1}{\sqrt{6}}\left\{|\uparrow^A \uparrow^B \downarrow^C\rangle + |\uparrow^A \downarrow^B \uparrow^C\rangle - 2|\downarrow^A \uparrow^B \uparrow^C\rangle\right\}$$

$$|D_1^-\rangle = \frac{1}{\sqrt{6}}\left\{|\downarrow^A \downarrow^B \uparrow^C\rangle + |\downarrow^A \uparrow^B \downarrow^C\rangle - 2|\uparrow^A \downarrow^B \downarrow^C\rangle\right\} \tag{17}$$

$$|D_2^+\rangle = \frac{1}{\sqrt{2}}\left\{|\uparrow^A \uparrow^B \downarrow^C\rangle - |\uparrow^A \downarrow^B \uparrow^C\rangle\right\}; \quad |D_2^-\rangle = \frac{1}{\sqrt{2}}\left\{|\downarrow^A \downarrow^B \uparrow^C\rangle - |\downarrow^A \uparrow^B \downarrow^C\rangle\right\}$$

$$\tag{18}$$

The Werner state formed out of the three particle state similar to eq.(13) leads to quantum entanglement if x>1/5 by means of the conditional Tsallis entropy method, also coinciding with the necessary and sufficient criterion for entanglement found by other independent methods. We now consider briefly the aspect of control of entanglement under the influence of environment as described by the Lindblad equation.

IV CONTROL OF ENTANGLEMENT

We now return to the question of what happens when one considers Lindblad evolution of the states, given their initial state leading to the notion of control. In [7, 8], oscillator systems were considered. Harmonic oscillators are prototypical of electronic nano-devices (L,C,R circuits). It is worth pointing out that in the case of bivariate

Gaussian distribution, for classical variables only the Schwarz inequality among the dispersions and correlations hold, whereas for the quantum variables, one has in addition the Heisenberg inequalities among the non-commuting variables (e.g., momentum and coordinate). The quantum entanglement is a reflection of this quantum correlations due to the uncertainty principle and thus is subtle. In the case of two-oscillator systems [7], these aspects are spelled out. Also, it was found that an initially unentangled pair may get entangled as time evolves and vice-versa exhibiting revivals. This is an interesting result in that it shows that one may be able to work in time zones where the system is entangled or, if that is needed, one can go to the region where they are not entangled! Interpreting the parameters of the Lindblad theory as environmental features in certain experimental situations gives us clues as to the possible control of the system behavior.

V. CONCLUDING REMARKS

In this work, we have presented the Lindblad dynamics of dissipative quantum systems and its implication for the dynamics of entangled quantum systems. This topic of research is of great interest in determining the feasibility of quantum device systems under the influence of dissipative environments. Thus the study of quantum phenomena under the action of Lindblad evolution is shown to be fundamental with practical consequence. The three topics considered here, dissipation, quantum entanglement, and their behaviour under the influence of dissipation, together promise to be of importance in the operation and construction of quantum nano-devices in the near future.

**ACKNOWLEDGEMENTS:** The authors are supported in part by the Office of Naval Research. They also thank Dr. Peter Reynolds of the Office of Naval Research for supporting this work.

TABLE I: Classifying Three Particle (A, B, C) Entangled States by Permutation symmetry (S, AS, NS), and Robustness or Fragility (R, r, or F).

| States | AB | | AC | | BC | | ABC |
|---|---|---|---|---|---|---|---|
| **GFF$^\pm$** $\equiv$ **GHZ$^\pm$** | S A,B,C | F | S A,B,C | F | S A,B,C | F | S A,B,C |
| **GFR$^\pm$** | NS | F | NS | F | AS B,C | R | AS B,C |
| **WRr$^\pm$** | NS | R | NS | R | S B,C | r | S B,C |
| **WRR$^\pm$** | S A,B,C | R | S A,B,C | R | S A,B,C | R | S A,B,C |